\documentclass[epjST]{svjour}
\usepackage{graphics}
\begin{document}
%
\title{A mean-field theory for self-propelled particles interacting by velocity alignment mechanisms}


\author{Fernando Peruani \inst{1}\inst{2} \and Andreas Deutsch\inst{2} \and Markus B\"ar\inst{3}}
\institute{ Max Planck Institute for
Physics of Complex Systems, N\"othnitzer Str. 38, 01187 Dresden,
Germany \\
\and Center for Information Services and High Performance Computing, Technische Universit\"at Dresden, Zellescher Weg 12, 01069 Dresden, Germany \\
\and Physikalisch-Technische Bundesanstalt, Abbestr. 2-12,
10587 Berlin, Germany.}
\abstract{A mean-field approach (MFA)  is proposed for the analysis of 
orientational order in  a two-dimensional system of stochastic self-propelled particles 
interacting by local velocity alignment mechanism.
The treatment is applied to  the cases of ferromagnetic (F) and liquid-crystal (LC) alignment. 
In both cases, MFA yields a second order phase transition for a critical noise strength and a scaling
exponent of 1/2 for the respective order parameters. 
We find that the critical noise amplitude $\eta_c$ at which orientational order emerges in the LC case
is smaller than in the F-alignment case, {\it i.e.} $\eta^{LC}_{C}<\eta^{F}_{C}$.
A comparison with simulations of individual-based models with F- resp. LC-alignment shows that the
predictions about the critical behavior and the qualitative relation between the respective critical
noise amplitudes are correct.} 

\maketitle
%


\section{Introduction}
Self-propelled particles (SPPs) with local interactions can organize into large-scale patterns. Flocks of
birds \cite{birds}, swarms of bacteria \cite{bacteria,myxobacteria}, sperm cells \cite{riedel_05}, mixtures of 
microtubules and motors  \cite{nedelec_97}, are examples for such a behavior. 
Beyond the complexity of each particular system, we observe that there are
some few common features which cause the emergence of long-range order in these systems:
the active motion of the particles and a velocity alignment mechanism.

The Vicsek-model \cite{vicsek95} is considered the simplest model of SPPs which
exhibits collective motion. In this model, point-like particles moving with a velocity vector of
constant magnitude interact by aligning their velocity direction to the local average velocity. 
One can think of this model as a model of moving spins, in which the velocity of the particles is
given by the spin-vector. Going further in this analogy with spin systems we denote this alignment
mechanism as ferromagnetic (F-alignment). 
The temperature associated with spin-systems enters in the Vicsek-model as noise in the alignment
mechanism. 
It was shown that two-dimensional SPPs with F-alignment and additive noise exhibit a second-order
phase transition which leads to low enough noise to long-range orientational order \cite{vicsek95,vicsek07,chate_comment}.
For different choices of system parameters and different choices of the noise term (non-additive noise term),
simulations show, however,  first order phase transitions to orientational order \cite{chate}. 
Interestingly, in equilibrium systems of non-moving spins with continuum symmetry this transition
cannot occur \cite{xy_model}.

F-alignment is one possible alignment mechanism, but clearly not the only one. 
If a system of self-propelled rods interacts simply by volume exclusion as described in \cite{peruani_06},
 particles may end up moving in the same direction as well as in opposite directions.
A biological realization of such a system are myxobacteria, which in the early stages of their life-cycle organize 
their motion by simply pushing each other \cite{myxobacteria}.
A similar effect without active motion occurs in liquid-crystals at high density where particles get locally 
 aligned  \cite{doi}. 
In analogy to these systems we name this  mechanism hence liquid-crystal alignment (LC-alignment). 
In a system of SPPs with LC-alignment particles align their velocity to the local average director. 
In  simulations of a  model of SPPs with LC-alignment it was found that at high density 
these moving "liquid-crystal" spins exhibit a second-order phase transition leading to 
long-range orientational order for low noise \cite{peruani_07}. 
Notice that the orientational order observed in SPPs with LC-alignment refers to the emergence of a 
global director in the system, while for F-alignment orientational order refers to the appearance of a global
 direction of motion.

Toner and Tu were the first to look for a macroscopic description of SPPs with F-alignment.
 Their approach was a phenomenological hydrodynamical description based on symmetry arguments for which
 they derived general macroscopic equations for a large class of  individual-based models of SPPs with F-alignment 
\cite{tu_95,toner_98,tu_98}. 
In this approach many of the parameters in the model are difficult to derive from the microscopic dynamics.
Recently Gr\'egoire et al. proposed an alternative approach based on the Boltzmann equation and could explain in a systematic way the functional form proposed by Toner and Tu \cite{gregoire_06} by use of an ad-hoc collision term. 

The case of LC-alignment has been much less explored. Recently, Ramaswamy et al. proposed a phenomenological hydrodynamical 
description for driven, but non-persistent, particles with LC-alignment \cite{ramaswamy_03}. 
The approach is comparable to the Toner and Tu description for F-alignment. 
One of the striking results of this approach is the giant number fluctuations of particles in the ordered state, 
which has been confirmed in simulations by Chat\'e et al. \cite{chate_06}.
More recent simulations have suggested that these fluctuations are linked with intrinsic phase separation of SPPs into
regions of high and low density \cite{ramaswamy_06}.

Here, a mean-field type description for SPPs with F- and LC- alignment is derived. 
Numerical evidence provided by individual-based simulations indicates that SPPs with
 both F and LC-alignment (and additive noise)  
can exhibit a continuous kinetic phase transition in two dimensions.
The derived mean-field equations  allow us to study ferromagnetic as well as liquid-crystal interactions among particles. 
Through this approach the phase transition to orientational order observed in individual-based simulations at high density for F and LC-alignment 
is correctly captured. 
In addition, we show that the critical noise amplitude $\eta_c$ is such that $\eta^{LC}_{C} < \eta^{F}_{C}$ in the mean-field
description as well as in the individual-based simulations in both investigated cases.

The paper is organized as follows. In section 2 we introduce an individual-based model of SPPs which interact
by either F- or LC-alignment, and give the corresponding definitions of the order parameters. 
In section 3 we present a macroscopic description of the individual-based model introduced in Section 2. We simplified the corresponding macroscopic model by considering a
mean-field ansatz, and
perform a linear stability analysis of the equations. In section 4 we compare the mean-field description with high density simulations in the limit of very fast angular relaxation. We discuss the limitations of the
mean-field approach in section 5.

%


\begin{figure}

\centering
\resizebox{0.9\columnwidth}{!}{
 \includegraphics{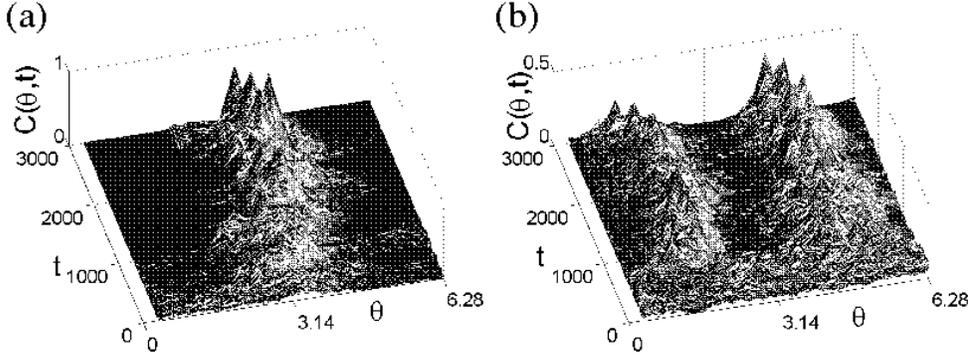} }

\caption{Temporal evolution of the velocity direction distribution (angular distribution) in simulations with very fast angular relaxation.
(a) corresponds to F-alignment, while (b) to LC-alignment. 
%
%
Number of particles $N=100$, radius of interaction $\epsilon=2$, 
linear system size $L=42.4$, and noise amplitude $\eta=0.25$.} \label{sim_veldistrib}
\end{figure}

\section{Individual-based model}

\subsection{Equations of motion}
We consider point-like particles moving at constant speed in a two
dimensional space and assume an over-damped situation such that the
state of particle $i$ at time $t$ is given by  its position $\mathbf{x}_i$ and its
direction of motion $\theta_i$. The evolution of these quantities follow:
\begin{eqnarray}\label{eq_mot_x}
\dot{\mathbf{x}}_i&=& v_0 \mathbf{v}(\theta_i)  \\
\label{eq_mot_angle} \dot{\theta_i}&=& - \gamma \frac{\partial
U}{\partial \theta_i}(\mathbf{x_i},\theta_i) +\tilde{\eta}_{i}(t)
\end{eqnarray}
where $\gamma$ is a relaxation constant, and $U$ the interaction potential between particles, and hence $\frac{\partial U}{\partial
\theta_i}(\mathbf{x_i},\theta_i)$ defines the velocity alignment mechanism. Moreover,  $v_0$
represents the active velocity of the particles, $\mathbf{v}(\theta_i)$ is defined as
$\mathbf{v}(\theta_i)=(\cos(\theta_i), \sin(\theta_i))$. The noise applied to the direction of motion, $\tilde{\eta}_{i}(t)$, obeys the following statistics: $\langle \tilde{\eta}_{i}(t) \rangle=0$ 
and $\langle \tilde{\eta}_{i}(t) \tilde{\eta}_{j}(t') \rangle=\tilde{\Gamma} \delta(t-t') \delta_{i,j}$, where $\delta(t-t')$ is a Dirac-delta function, $\delta_{i,j}$ is a
Kronecker-delta function, and $\tilde{\Gamma}$ is the "strength" of the noise.
The evolution Eqs. (\ref{eq_mot_x}) and (\ref{eq_mot_angle})
are expressed in terms of first derivatives. In this way, $v_0$ in Eq. (\ref{eq_mot_x}) can be
considered as an active force divided by a translational friction
coefficient, and  $\gamma$ in Eq. (\ref{eq_mot_angle}) as the inverse of a
rotational friction coefficient.

In analogy to spin systems, the ferromagnetic velocity alignment
mechanism is given by a potential defined as:
\begin{equation} \label{eq:ferromagneticpotential}
U_{F}(\mathbf{x_i},\theta_i)=-\sum_{\left|\mathbf{x}_{i}-\mathbf{x}_{j}\right|\leq\epsilon}
\cos(\theta_i-\theta_j)
\end{equation}
where $\epsilon$ is the radius of
interaction of the particles. For the liquid-crystal alignment mechanism, we choose the
potential introduced by Lebwohl and Lasher to study liquid
crystal interactions on a lattice \cite{lebwohl} which reads:
\begin{equation} \label{eq:liquidcrystalpotential}
U_{LC}(\mathbf{x_i},\theta_i)=-\sum_{\left|\mathbf{x}_{i}-\mathbf{x}_{j}\right|\leq\epsilon}
\cos^2(\theta_i-\theta_j)
\end{equation}
One can add a coupling strength coefficient to the expression (\ref{eq:ferromagneticpotential}) and (\ref{eq:liquidcrystalpotential}). We assume that the
coupling strength is absorbed in $\gamma$ in Eq. (\ref{eq_mot_angle}).
Notice that the potential given by Eq. (\ref{eq:ferromagneticpotential}) exhibits one minimum, while Eq. (\ref{eq:liquidcrystalpotential}) has two minima,
which correspond to particles pointing in the same direction and particles pointing in opposite directions.

In the limiting case of very fast angular relaxation we obtain 
from Eqs. (\ref{eq_mot_x}) and (\ref{eq_mot_angle}) the updating rules:
\begin{eqnarray}\label{motion_pos}
\mathbf{x}_{i}^{t+\Delta t }&=&\mathbf{x}_{i}^{t} +v_0 \mathbf{v}\left(\theta_i^{t} \right)\Delta t \\
\label{motion_vel} \theta_i^{t+\Delta t }
&=&\arg\left(\sum_{\left|\mathbf{x}_{i}^{t}-\mathbf{x}_{j}^{t}\right|\leq\epsilon}\mathbf{f}(\mathbf{v}(\theta_j^{t}),\mathbf{v}(\theta_i^{t}))\right)+\eta_{i}^{t}
\end{eqnarray}
where  $\arg\left(\mathbf{b}\right)$
indicates the angle of a vector $\mathbf{b}$ in polar coordinates, and  
$\eta_{i}^{t}$ is obtained from a distribution $p(\eta_{i}^{t})$ defined as $p(\eta_{i}^{t})=1/\eta$ when 
$\eta_{i}^{t}$ belongs to the interval $\left[-\frac{\eta}{2},\frac{\eta}{2}\right]$, and $p(\eta_{i}^{t})=0$ otherwise. 
In consequence, $\langle \eta_{i}^{t} \rangle=0$ and $\langle \eta_{i}^{t} \eta_{j}^{t'} \rangle=(\eta^2/12) \delta_{i,j} \delta_{t,t'}$. 
Given two vectors $\mathbf{a}$ and $\mathbf{b}$, the function $\mathbf{f}(\mathbf{a},\mathbf{b})$ is defined as follows. 
For F-alignment, $\mathbf{f}(\mathbf{a},\mathbf{b})=\mathbf{a}$.
For LC-alignment, $\mathbf{f}$ takes the form:
\begin{equation} \label{function_f}
\mathbf{f}\left(\mathbf{a},\mathbf{b}\right) = \left\{
\begin{array}{lcr}
\mathbf{a} & \mbox{if} & \mathbf{a}.\mathbf{b} \geq 0 \\
-\mathbf{a} & \mbox{if} & \mathbf{a}.\mathbf{b}<0
\end{array} \right.
\end{equation}
as described in \cite{peruani_07}. Notice that F-alignment implies that the sum in Eq. (\ref{motion_vel}) becomes a simple weighted local velocity average. 
In consequence, under F-alignment Eqs. (\ref{motion_pos}) and (\ref{motion_vel}) define the original Vicsek model \cite{vicsek95}. 
Particles interact by calculating the local average direction of motion, and if the noise strength is low enough move roughly in that direction. 
In contrast, LC-alignment implies a local average of mapped velocities that leads particles to calculate the local average director (and not the average direction
of motion as for F-alignment). This process defines locally two possible directions of motion, and particles choose from these two options the one that is closer
to their present direction of motion. 
Notice that throughout the text F-alignment refers to the Vicsek model, and this can be either in its original discrete-time form, 
or in its generalized continuum time form given by Eq. (\ref{eq_mot_x}), (\ref{eq_mot_angle}), and (\ref{eq:ferromagneticpotential}).

\subsection{Order parameters}
If particles interact through the F-alignment mechanism, and assuming low noise
amplitude, they get locally aligned, and locally point in a similar direction. 
The question is whether such local alignment may lead to a global orientational order
in which a macroscopic fraction of the particles in the system points in a similar direction. 
The order parameter that quantifies this phenomenon is the modulus of the normalized total momentum (analogous
to the magnetization in the XY-model\cite{xy_model}) that we express as:
\begin{eqnarray}\label{eq:orderparam_f}
S^{F} = \left| \frac{1}{N} \sum_{i=0}^{N} \mathbf{v}\left(\theta_i^{t} \right) \right|
\end{eqnarray}
where $N$ stands for the total number of particles in the system. $S^{F}$ takes the
value $1$ when all particle move in the same direction. On the other hand, $S^{F}$
is equal to $0$ in the disordered case in which particles point in any direction
with equal probability. 
This can be also observed through the velocity direction distribution, that in two dimensions becomes an angular distribution $C(\theta)$. 
For high values of the noise, $C(\theta)$ is flat. 
When the noise is decreased below a critical noise $\eta_c$ an instability arises in the system (characterized by a single peak) indicating the onset of orientational order as shown in Fig. \ref{sim_veldistrib}(a). 

On the other hand, if for example, half of the particles move in one
direction, and the other half in the opposite direction, $S^{F}$ is also $0$.
Clearly, $S^{F}$  cannot distinguish such a state and the completely disordered state. 
However, LC-alignment may induce such a kind of local arrangement of particle
velocities, and lead to a global orientational order state
in which there are two opposite main directions of motion in the system. 
To study such orientation ordering, one uses the order matrix $Q$ of
liquid crystals \cite{doi}. For two dimensions one takes the
largest eigenvalue $S^{LC}$ of $Q$ and obtain the following scalar orientational
order parameter:
\begin{equation}\label{eq:orderparam_lc}
S^{LC}=\frac{1}{4}+\frac{3}{2}\sqrt{\frac{1}{4}-\frac{1}{N^{2}}\left\{
\sum_{i,j}^{N}v_{xi}^{2}v_{yj}^{2}-v_{xi}v_{yi}v_{xj}v_{yj}\right\}
}
\end{equation}
where $v_{xi}$ and $v_{yi}$ are defined as $v_{xi}=\cos(\theta_i)$
and $v_{yi}=\sin(\theta_i)$.  The orientational order parameter $S^{LC}$
takes the value $1$ when all particles are aligned along
the same director, and the value $\frac{1}{4}$ in the disordered phase 
where particles move with equal probability in any direction.
Again this can be observed through the velocity direction distribution $C(\theta)$. 
In this case, for low values of the noise amplitude, as shown in Fig. \ref{sim_veldistrib}(b), an instability arises in the system with the characteristic of having two peaks separated by $2 \pi$.


\section{Mean-field approach}

\subsection{Derivation of the mean-field approach}

A system of SPPs may alternatively be described through a density field
$\psi(\mathbf{x},\mathbf{v}(\theta),t)=\psi(\mathbf{x},\theta,t)$ in such a way that  
 the particle density at a point $\mathbf{x}$ is given by
\begin{equation}\label{eq:density}
\rho \left(\mathbf{x},t\right)=\int_{0}^{2 \pi}\psi\left(\mathbf{x},\theta,t\right)d\theta
\end{equation}
while the velocity direction distribution (or angular distribution) can be expressed
as:
\begin{equation}\label{eq:angulardistribution}
C \left(\theta,t\right)=\int_{\Omega} \psi\left(\mathbf{x},\theta,t\right){d\mathbf{x}}
\end{equation}

We recall that in the
individual-based model the kinetic energy is conserved, while 
the momentum is not. For 
F-alignment, the system tends to increase the total
momentum, while for LC-alignment the tendency is to
decrease it. The continuum approach has to reflect that particles
can re-orient their velocity direction but always move at constant speed. On the other hand, the number of particles has to be
conserved. 
%
%
%
Under these assumptions the following evolution
equation for $\psi(\mathbf{x},\theta,t)$ is obtained:
\begin{equation}\label{eq_pde_evolution}
\partial_{t} \psi = D_{\theta}\partial_{\theta\theta}\psi-\partial_{\theta}\left[F_{\theta}\psi\right]- \mathbf{\bigtriangledown} \left[\mathbf{F}_{x}\psi\right]
\end{equation}
where $F_{\theta}\psi$ and $\mathbf{F}_{x}\psi$ are deterministic fluxes
which are associated to the local alignment mechanism and active
migration, respectively,  and $D_{\theta}$ refers to the diffusion
in the direction of motion. 

Let us derive the specific expressions for $D_{\theta}$, $F_{\theta}$ and
$\mathbf{F}_{x}$. 
$D_{\theta}$ depends on the square of the noise amplitude. For example, in the individual-based simulations
$\eta_{i}(t)$  has been taken, as mentioned above, from a homogeneous distribution of width $\eta$ and centered around $0$, and in consequence $D_{\theta}$ is given  $D_{\theta} = \eta^2 \Delta t/24 $, where  $\Delta t$ is the temporal time step.
$F_{\theta}$ contains the interaction of a
particle located at $\mathbf{x}$ and pointing in direction $\theta$
with all neighboring particles which are at a distance less than
$\epsilon$ from $\mathbf{x}$, and so takes the form:
\begin{equation}\label{eq_f_theta}
F_{\theta}=-\gamma \int_{R(\mathbf{x})}{d\mathbf{x}'}\int^{2 \pi}_{0}{d \theta'} \frac{\partial U(\mathbf{x},\theta,\mathbf{x}',\theta')}{\partial \theta} \psi(x',\theta',t)
\end{equation}
where $U(\mathbf{x},\theta,\mathbf{x}',\theta')$ represents the pair
potential between a particle located at $\mathbf{x}$ and pointing in
direction $\theta$, and another at $\mathbf{x'}$ and pointing in
direction $\theta'$. $R(\mathbf{x})$ denotes the interaction
neighborhood around $\mathbf{x}$. If $\mathbf{x}'$ is inside $R(\mathbf{x})$, then
$U(\mathbf{x},\theta,\mathbf{x}',\theta')=U(\theta,\theta')$. 
%
%
Finally, $F_{\theta}$ can be thought as the "torque" felt by a particle
located at $\mathbf{x}$ and pointing in direction $\theta$.
The expression for $\mathbf{F}_{x}$ is straightforward and is
directly related to the velocity of particle at $\mathbf{x}$ and
pointing in direction $\theta$, 
\begin{equation}\label{eq_f_x}
\mathbf{F}_{x}=v_0\mathbf{v}(\theta)
\end{equation}
%


\begin{figure}

\centering

\resizebox{0.7\columnwidth}{!}{
 \includegraphics{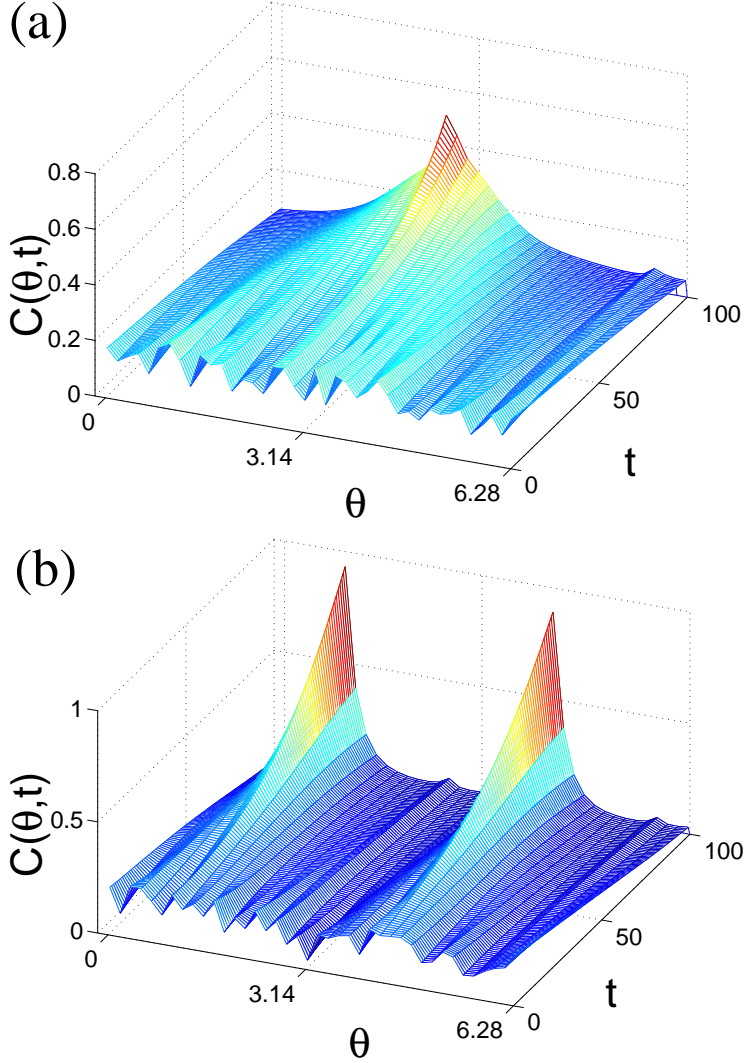} }

\caption{Temporal evolution of $C(\theta,t)$. (a) F-alignment, numerical integration of Eq. (\ref{eq:parallel}) with $D_{\theta}=0.28$. 
(b) LC-alignment, numerical integration of Eq. (\ref{eq:parallelantiparallel}) with $D_{\theta}=0.014$. 
For both (a) and (b), $C^{*}=0.3183$, $\Delta t = 0.001$ and $\Delta
\theta = 0.16$. The initial condition is a random perturbation around $C^{*}$.
Notice that for F-alignment a single peak emerges, while for LC-alignment the distribution develops two peaks.}
\label{num_example_angle}
\end{figure}

\subsection{Angular distribution}

Integrating both sides of Eq. (\ref{eq_pde_evolution}) over the space $\Omega$ we
obtain an evolution equation for $C(\theta, t)$ which still depends on
$\psi(\mathbf{x},\theta,t)$. In the following we assume a homogeneous spatial
distribution of particles $\psi(\mathbf{x},\theta,t)=C(\theta,t) \rho_0/N$, where $\rho_0$ is defined as 
$\rho_0=N/L^2$, being $L$ the linear size of the system. With these assumptions the
equation for the temporal evolution of $C(\theta, t)$ reads:

\begin{equation}\label{eq:angulardistribution}
\frac{\partial C(\theta, t)}{\partial
t}=D_{\theta}\frac{\partial^2 C(\theta,t)}{\partial \theta^2}+\gamma \frac{\pi
\epsilon^2}{L^2} \partial_{\theta}\left[ \left\{\int_{0}^{2 \pi} {d\theta'}
\frac{\partial U(\theta,\theta')}{\partial \theta} C(\theta',t)  \right\} C(\theta,t) \right]
\end{equation}

\subsection{Linear stability analysis for F-alignment}

For both F- and LC-alignment the homogeneous angular distribution is a steady state of Eq.
(\ref{eq:angulardistribution}). We determine the onset of the ordered state by studying the linear
stability of the disordered state. First let us look at the F-alignment. By dividing both sides of Eq.
(\ref{eq:angulardistribution}) by $\gamma \pi \epsilon^2 /L^2$, and redefining time as $\tau = (\gamma \pi
\epsilon^2 /L^2) t$, and $D_{\theta}'= D_{\theta}/ [\gamma \pi \epsilon^2 /L^2]$ one obtains:
\begin{equation}\label{eq:parallel}
\frac{\partial C\left(\theta,t\right)}{\partial
\tau}=D_{\theta}'\partial_{\theta\theta}C\left(\theta,t\right)+\partial_{\theta}\left[\left\{
\int
d\theta'\sin\left(\theta-\theta'\right)C\left(\theta',t\right)\right\}
C\left(\theta,t\right)\right]
\end{equation}

Now, consider a weak perturbation of the homogeneous pattern:
\begin{equation}\label{eq:perturb_ang}
C\left(\theta,t\right)=C^{*}+C_{0}e^{in\theta}e^{\lambda \tau}
\end{equation}
Notice that ${e^{in\theta}}$ are eigenfunctions of 
the operators emerging from the linearization of Eq. (\ref{eq:parallel}) about the homogeneous steady state. By substituting
 into Eq. (\ref{eq:parallel}) and keeping terms linear in $C_{0}$
we obtain the following expression for the eigenvalues:
\begin{equation}\label{eq:eigenvalue_parallel}
Re(\lambda)=-D_{\theta}' n^{2}+\pi C^{*}\delta_{n,1}
\end{equation}
This means that the only mode which can become unstable is
$n=1$. The condition for the instability of the homogeneous state
takes the form:
\begin{equation}\label{eq:criticaldensity_ferro}
\rho_0>\frac{2 D_{\theta}}{\gamma \pi \epsilon^2}
\end{equation}
where $\rho_0 = N/L^2$. For a given noise amplitude, expressed by $D_{\theta}$, there is a
critical particle density above which the homogeneous solution is
no longer stable.
Fig. \ref{sim_veldistrib}(a) shows that in individual-based 
simulations indeed a single peak emerges in the system for low density.
Fig. \ref{num_example_angle}(a) confirms that such qualitative behavior is
recovered by numerical integration of Eq.(\ref{eq:parallel}).

\subsection{Linear stability analysis for LC-alignment}

Applying analogous procedure for LC-alignment yields:
\begin{equation}\label{eq:parallelantiparallel}
\frac{\partial C\left(\theta,t\right)}{\partial
\tau}=D_{\theta}'\partial_{\theta\theta}C\left(\theta,t\right)+\partial_{\theta}\left[\left\{
\int
d\theta' 2\cos\left(\theta-\theta'\right)\sin\left(\theta-\theta'\right) C\left(\theta',t\right)\right\}
C\left(\theta,t\right)\right]
\end{equation}
Again the weakly perturbed homogeneous ansatz given by Eq. (\ref{eq:perturb_ang}) is considered. As before $e^{in\theta}$ are eigenfunctions of the linearized operators. Substituting Eq. (\ref{eq:perturb_ang}) into Eq.
(\ref{eq:parallelantiparallel}) and keeping terms linear in $C_{0}$
the following expression for the eigenvalues is obtained:
\begin{equation}\label{eq:eigenvalue_parallel}
Re(\lambda)=-D_{\theta}' n^{2}+2 \pi C^{*} \delta_{n,2}
\end{equation}
As for the F-alignment, there is only one mode which could become unstable, but this time it is $n=2$. This mode is the only one that exhibits two peaks
separated by $\pi$, which corresponds to two population of particles migrating in opposite direction.
In this case, the instability condition of the homogeneous states takes simply the form:
\begin{equation}\label{eq:criticaldensity_LC}
\rho_0>\frac{4 D_{\theta}}{\gamma \pi \epsilon^2}
\end{equation}
Again, this inequality defines a critical density for a given
noise amplitude above which the homogeneous solution is no longer
stable. Fig. \ref{sim_veldistrib}(b) shows the emergence of these two peaks for LC-alignment in individual-based 
simulations. Numerical integration of Eq. (\ref{eq:parallelantiparallel}), see Fig. \ref{num_example_angle}(b), confirms that this behavior is
recovered qualitatively by the mean-field description.

Eqs. (\ref{eq:criticaldensity_ferro}) and (\ref{eq:criticaldensity_LC}) indicate that the instability of the homogeneous
state is given by $\rho_0$, $D_{\theta}$, and $\epsilon$, the range of interaction. The
critical density is inversely proportional to $\epsilon^2$, hence when $\epsilon$ goes to infinity the critical density
goes to $0$. The interpretation of this is straightforward,
$\epsilon\longrightarrow\infty$ indicates that particles have 
infinity "visibility", i.e., each particle can sense all other 
particles in the system. In this way, the collective behavior
has to emerge independent of particle density. The other limiting
case is represented by $\epsilon\longrightarrow0$. In this case particles do not interact 
 and in consequence no organized motion is possible.

\begin{figure}

\centering

\resizebox{0.6\columnwidth}{!}{
 \includegraphics{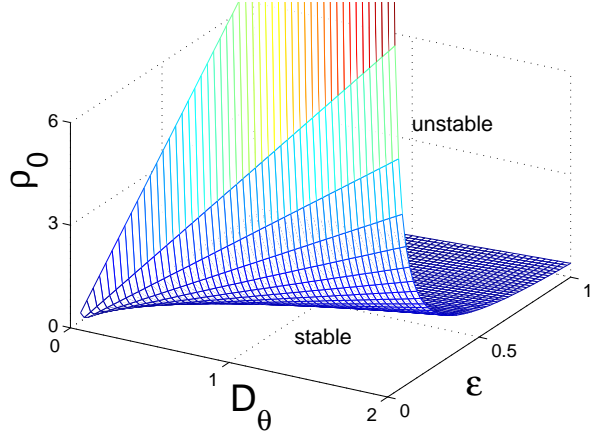} }

\caption{Phase diagram derived from the continuum approach. The
unstable region corresponds to the velocity orientational order,
while stable means no orientational order can be observed.} \label{phasediagram}
\end{figure}

From these findings a phase diagram is derived that shows where
the system exhibits velocity orientational order (see Fig. \ref{phasediagram}).

\subsection{Spatially inhomogeneous steady states}

\begin{figure}

\centering
\resizebox{0.9\columnwidth}{!}{
 \includegraphics{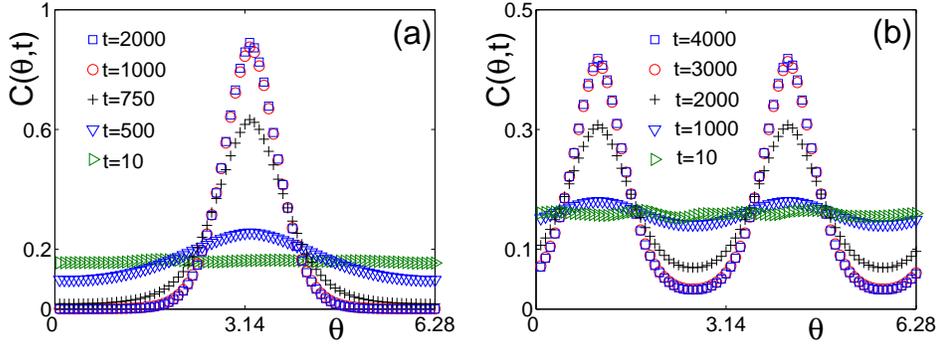} }

\caption{Convergence towards the non-trivial stable steady state. (a) F-alignment, numerical integration of Eq. (\ref{eq:parallel}) with $D_{\theta}=0.3472$. 
(b) LC-alignment, numerical integration of Eq. (\ref{eq:parallelantiparallel}) with $D_{\theta}=0.2813$. For both (a) and (b), $C^{*}=0.3183$, $\Delta t = 0.001$ and $\Delta
\theta = 0.0785$. The initial condition is a random perturbation around $C^{*}$. 
Different curves correspond to different times. Notice that for large values of $t$ curves start to overlap on top of each other.}
\label{fig_steadystates}
\end{figure}

Through the linear stability analysis it has been found for which conditions the homogeneous distribution (disordered state) 
becomes unstable. 
To study the nonlinear behavior of these instabilities in more detail, Eq. (\ref{eq:angulardistribution}) can be integrated numerically.
Details about the numerical methods  are given in the Appendix.
Fig. \ref{num_example_angle} has shown already the temporal evolution of $C(\theta,t)$. The initial condition is a homogeneous state with small random perturbations: $C(\theta_n,
t=0)= C^{*} + \eta(n)$, where $\theta_n$ denotes the discrete angular variable, $C^{*}$ is the constant unperturbed homogeneous state, which we have set to be in the unstable regime
 according to Eqs. (\ref{eq:criticaldensity_ferro}) and (\ref{eq:criticaldensity_LC}) for the F- and LC-alignment case, respectively, and $\eta(n)$ is a white
 noise.
In Fig. \ref{fig_steadystates} the angular distribution for F- and LC-alignment is shown at different times. $C(\theta,t)$ tends asymptotically to a
non-trivial steady state, see Fig. \ref{fig_steadystates}.
The width of the peaks in the steady state is the result of the competition between influence of rotational diffusion, indicated by $D_{\theta}$, and the alignment force  associated with the interactions.

\subsection{Scaling of the order parameter close to the transition}

For a given density, there is a critical $D_{\theta_{c}}$. Close to $D_{\theta_{c}}$ we expect to observe that only one mode dominates $C(\theta,t)$. As said before, $n=1$ is
dominant for F-alignment and $n=2$ governs LC-alignment. 
The steady state distribution $C_{st}(\theta)$ then takes the form:
\begin{eqnarray}\label{eq:steady_f}
C_{st}(\theta) \simeq C^{*} + B_{1} \sqrt{D_{\theta_{c}} - D_{\theta}} \cos(\theta - \theta_0)
\end{eqnarray}
for F-alignment,  while for LC-alignment the expression reads:
\begin{eqnarray}\label{eq:steady_lc}
C_{st}(\theta)\simeq C^{*} + B_{2} \sqrt{D_{\theta_{c}} - D_{\theta}} \cos(2 (\theta - \theta_0))
\end{eqnarray}
where $B_{1}, B_{1}$ are constants and $\theta_0$ is an arbitrary phase which depends on the initial condition.
In both cases the maximum amplitude of $C_{st}(\theta)$ close to the $D_{\theta_{c}}$ grows as $\sqrt{D_{\theta_{c}} - D_{\theta}}$. Inserting Eq. (\ref{eq:steady_f}) into
Eq. (\ref{eq:orderparam_f}) and using, as indicated above, $D_{\theta}=\eta^2 \Delta t/24$, we obtain the scaling of the order parameter $S^{F}$:
\begin{eqnarray}\label{eq:scaling_f}
S^{F} \simeq \tilde{B}_{1} \sqrt{\eta_c - \eta}
\end{eqnarray}
where $\tilde{B}$ is a constant. To obtain the scaling of the order parameter $S^{LC}$, we insert (\ref{eq:steady_lc}) into 
Eq. (\ref{eq:orderparam_lc}):
\begin{eqnarray}\label{eq:scaling_lc}
S^{LC} \simeq \frac{1}{4} + \tilde{B}_{2} \sqrt{\eta_c - \eta}
\end{eqnarray}
where again $\tilde{B}_{2}$ is a constant. $\tilde{B}_{1}$ and $\tilde{B}_{2}$ are constants proportional to $\eta_c$.


\section{Comparison with individual-based simulations}

Individual-based simulations have been performed in the limit case of very fast angular relaxation \cite{vicsek95,vicsek07}.
In contrast, our mean-field description  assumes that there is a finite angular relaxation. 
Can we expect the mean-field approach to describe scaling of the orientational dynamics in this kind of simulations?
We redefine $\gamma$ as function of
the particle velocity $v_0$ and the particle density $\rho$. The effective resulting mean-field equation reads:
\begin{equation}\label{eq:effective_meanfield}
\frac{\partial C}{\partial
t}(\theta, t)=D_{\theta}\frac{\partial^2 C(\theta,t)}{\partial \theta^2}+\gamma(v_0, \rho)  \frac{\partial}{\partial_{\theta}} \left[ \left\{\int_{0}^{2 \pi} {d\theta'}
\frac{\partial U(\theta,\theta')}{\partial \theta} C(\theta',t)  \right\} C(\theta,t) \right]
\end{equation}
where $\gamma(v_0, \rho)$ is an effective interaction strength which
absorbs the spatial dynamics.

The scaling obtained from individual-based simulations may now be compared with the one predicted by the mean-field approach. We recall that $D_{\theta} \sim \eta^{2}$, where $\eta$
is the orientational noise amplitude used in individual-based simulations. From this we find that $D_{\theta_{c}} - D_{\theta}$ has to be $D_{\theta_{c}} - D_{\theta} = K
(\eta_c - \eta) + O((\eta_c - \eta)^2)$, where $K$ is a constant. We focus on the LC-alignment and replace this expression into Eqs.
(\ref{eq:steady_lc}) and (\ref{eq:scaling_lc}). We obtain that $C_{st}(\theta)\simeq C^{*} + B_1 \sqrt{\eta_{c} - \eta} \cos(2 (\theta - \theta_0))$ and $S^{LC} \simeq
\frac{1}{4} + B_2 \sqrt{\eta_c - \eta}$, where $B_1$ and $B_2$ are constants. 

\begin{figure}

\centering

\resizebox{0.85\columnwidth}{!}{
 \includegraphics{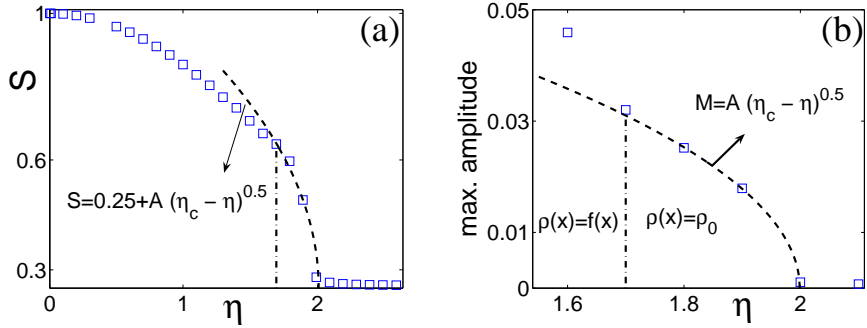} }

\caption{(a) Scaling of the scalar order parameter $S^{LC}$ 
with the noise amplitude $\eta$. (b) Scaling of the maxima in $C_{st}(\theta)$ with $\eta$. 
Symbols correspond to simulations with LC-alignment in the limit of very fast angular relaxation. $\rho=4$ and $N=2^{12}$. 
The dashed-curve corresponds to the scaling predicted by the mean-field approach (see Eq. (\ref{eq:scaling_lc})). 
The vertical dot-dashed line indicates the onset of clustering effects in the simulations.
To the right of that line particle density $\rho(x)$ can be considered constant. To the left of the line, $\rho(x)$ becomes a function of $x$. 
For F-alignment in limit of very fast angular relaxation, the scaling of $S^{F}$  is close to the one predicted by Eq. (\ref{eq:scaling_f}), 
compare also \cite{vicsek95,vicsek07}.} \label{scaling_s_eta}
\end{figure}

Fig. \ref{scaling_s_eta}(a) shows a comparison between the scaling predicted by the mean-field approach for $S^{LC}$ (dashed curve) and the one obtained from individual-based simulations
for $\rho = 4$ in the limit of very fast angular relaxation (symbols). We find good agreement between the mean-field prediction and the simulations for the scaling of $S$ near $\eta_c$ that suggests that individual-based simulations with LC-alignment at high densities exhibit a
mean-field type transition. 
Notice that simulations start to deviate exactly at the point where density fluctuations become important (denoted by the dot-dashed vertical line in
(\ref{scaling_s_eta})(a)). Let us recall that the mean-field approach implies the assumption of homogeneous density.
Evidence also points towards a mean-field transition if we look at the scaling of the maximum amplitude of the angle distribution as function of the angular noise intensity
$\eta$ (see Fig. (\ref{scaling_s_eta})(b)). 
The order parameter scaling exponent for F-alignment (Vicsek-model) has been found to be 0.45 $\pm$ 0.07 \cite{vicsek95,vicsek07}, which is also in line 
with the predictions of the presented mean-field theory. 

\begin{figure}

\centering

\resizebox{0.55\columnwidth}{!}{
 \includegraphics{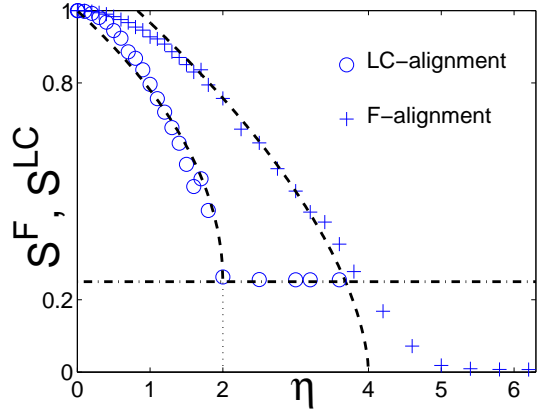} }

\caption{Comparing simulations of particles with F-alignment  
(crosses) and LC-alignment (circles) in the limiting case of very fast angular relaxation. In both cases
$N=2^{14}$ and $\rho=2.0$. Notice that the order parameter for F-alignment is $S^{F}$ while for LC-alignment is $S^{LC}$ (See text).
%
The dashed horizontal line indicates the minimum value that $S^{LC}$ could take. 
The dashed curves correspond to the best fit assuming an exponent $0.5$, i.e., $\eta_c$ was the fitting parameter.}
\label{fig_factor_2}
\end{figure}

Finally, Fig. \ref{fig_factor_2} shows that in individual-based simulations with the same parameters and different
(namely LC- and F-) alignment mechanism, in the limit of very fast angular relaxation $\eta^{LC}_{C}<\eta^{F}_{C}$ 
as predicted by the mean-field theory. 
Note, however, that the simulations yield  $2 \eta^{LC}_{C} \approx  \eta^{F}_{C}$, while the 
mean-field description predicts $\sqrt{2}\eta^{LC}_{C} = \eta^{F}_{C}$. 
%

\section{Concluding remarks}

We have derived a mean-field theory for self-propelled particles which accounts for F- and LC-alignment. 
This approach predicts a continuous phase transition with the order parameter scaling with an exponent one half in both cases.
In addition, it yields that  the critical noise amplitude below which orientational order emerges is smaller for LC-alignment 
than for F-alignment, i.e., $\eta^{LC}_{C}<\eta^{F}_{C}$.

These findings were confirmed by individual-based simulations with F- and LC-alignment. 
In the limit of infinitely fast angular relaxation used in simulations here the mean-field theory provides a good
 qualitative description of the simulations.
If simulations were performed by integrating Eqs. (\ref{eq_mot_x}) and (\ref{eq_mot_angle}) with a finite angular relaxation, 
{\it i. e.} a finite $\gamma$, a direct correspondence between parameters used in simulations and parameters in the mean-field theory
can be made and quantitative comparisons become possible. 
%
%
The presented mean-field theory is not an exact coarse-grained description of Eqs. (\ref{eq_mot_x}) and (\ref{eq_mot_angle}). 
For instance, we have neglected the potential impact of particle-particle correlations. 
Furthermore, we have assumed spatial homogeneous density to study the emergence of orientational order. 
Thus, the presented approach does not apply to situations where self-propelled particles show clustering at the onset of
orientational order \cite{peruani_06,peruani_07}. 
In summary, a better understanding of the problem should imply the study of the interplay between local orientational order and density fluctuations. We leave that for future
research.\\

{\bf Acknowledgments} We would like to thank E.M. Nicola and L.G. Morelli for fruitful discussions. This work
was financially supported by the Deutsche Forschungsgemeinschaft (DFG) through Grant No. DE842/2. 
F.P. and M.B. thank Yasumasa Nishiura and Hokkaido University for hospitality and generous financial support.


\section{Appendix - numerical integration scheme} \label{appe_a}


The numerical integration of the integro-partial differential
equation  (\ref{eq:angulardistribution}) requires to perform the numerical
integration of Eq. (\ref{eq_f_theta}) to then proceed to the
integration of the diffusive and advective terms in Eq. (\ref{eq:angulardistribution}).

At each time step $F_{\Omega}(\theta, t)$ is calculated through a
simple Newton-Cotes method. Then the integration of Eq. (\ref{eq:angulardistribution}) 
is performed through an operator splitting method. The diffusion is
implemented by an explicit forward method. The integration of the
active turning (Eq. (\ref{eq_f_theta})) contained in the advective term requires special
attention. Since $F_{\Omega}$ depends explicitly on $\theta$ and
$t$ neither a Lax  nor an Upwind method gives a satisfactory
result. We overcame this difficulty by implementing the following
variant of the Upwind method:

\begin{eqnarray}
\nonumber C(\theta_k, t_j+1) =
(1-|\widetilde{F}_{\Omega}(\theta_k, t_j)|)C(\theta_k,
t_j) + \\
\nonumber \Theta(\widetilde{F}_{\Omega}(\theta_k-1,
t_j))|\widetilde{F}_{\Omega}(\theta_k-1, t_j)|C(\theta_k-1, t_j)+
\\
 \Theta(-\widetilde{F}_{\Omega}(\theta_k+1, t_j))|\widetilde{F}_{\Omega}(\theta_k+1,
t_j)|C(\theta_k+1, t_j)
\end{eqnarray}

where $\theta_k$ and  $t_j$ represent the discrete indices of the
angular and temporal variables  respectively, $\Theta(x)$ denotes
a Heaviside function, and $\widetilde{F}_{\Omega}(\theta_k, t_j)$
is defined as $\widetilde{F}_{\Omega}(\theta_k, t_j)=(\Delta t /
\Delta \theta)F_{\Omega}(\theta_k, t_j)$ where $\Delta \theta$ and $\Delta
t$ are the discretization of the space and time respectively.


\begin{thebibliography}{99}
\bibitem{chate_comment}
 In \cite{chate} authors claim that even with additive angular noise the phase transition is of first-order.
\end{thebibliography}
\end{document}